\documentclass[12pt]{article}

\usepackage{amssymb}
\usepackage{amsmath}
\usepackage{color}
\usepackage{cite}

\makeatletter
 \def\eqalign#1{\null\vcenter{\def\\{\cr}\openup\jot\m@th
  \ialign{\strut$\displaystyle{##}$\hfil&$\displaystyle{{}##}$\hfil
      \crcr#1\crcr}}\,}
\makeatother

\def\ct{ \cite }

\def\p{ \partial }

\def\eq{\equiv}

\def\q{\quad}

\def\equ{Eq.~\eqref}

\title{Variational principle for the relativistic hydrodynamic flows with discontinuities, and local invariants of motion.}

\author{A.V. KATS \\
Usikov Institute for Radiophysics and Electronics \\ National
Academy of Sciences of Ukraine, \\
  61085, 12 Ak. Proskury St., Kharkiv, Ukraine, \\
e-mail: avkats@online.kharkiv.com\\
\vspace{0.5cm}\\
J. JUUL RASMUSSEN \\
Optics and Plasma Research Department, OPL-128,\\
 Ris{\o} National Laboratory, P.O. Box 49,\\
DK-4000 Roskilde, Denmark, \\
e-mail: jens.juul.rasmussen@risoe.dk}

\begin{document}

\maketitle


\begin{abstract}
A rigorous method for introducing the variational principle
describing relativistic ideal hydrodynamic  flows with all possible
types of discontinuities (including shocks) is presented in the
framework of an exact Clebsch type representation of the
four-velocity field as a bilinear combination of the scalar fields.
The boundary conditions for these fields on the discontinuities are
found. We also discuss the local invariants caused by the relabeling
symmetry of the problem and derive recursion relations linking
invariants of different types. These invariants are of specific
interest for stability problems. In particular, we present a set of
invariants based on the relativistic generalization of the Ertel
invariant.
\end{abstract}




\paragraph{Introduction.}


In this paper we discuss some problems related to ideal relativistic
hydrodynamic (RHD) flows in the framework of the special relativity.
They are pertinent to the description of flows with discontinuities,
including shocks, in terms of canonical (Hamiltonian) variables
based upon the corresponding variational principle and introducing
local invariants along with recursion relations. These subjects are
of interest from a general point of view and are very useful in
solving nonlinear problems, specifically, nonlinear stability
investigation, description of the turbulent flows, etc. In
particular, the use of the Hamiltonian approach along with
additional local invariants of the motion and the corresponding
Casimirs allows to improve the nonlinear stability criteria. The
necessity to consider the relativistic flows is motivated by a wide
area of applications, including the astrophysical and cosmological
problems.



Variational principles for the ideal relativistic hydrodynamic (RHD)
flows, without discontinuities, have been widely discussed in the
literature, see, for instance,
\cite{Schutz70,Brown93,ZakharovKuznetsov97} and citations therein.
As for the nonrelativistic flows, the least action principle is
conveniently formulated in terms of the subsidiary fields and the
corresponding velocity representation known as the Clebsch
representation, see
\cite{lamb,GP93,Berd_83,Serrin59,Lin_63,Salmon82}. These subsidiary
fields can be introduced explicitly by means of the Weber
transformation, \cite{Weber1868}, see also
\cite{lamb,ZakharovKuznetsov97}. Alternatively, they naturally arise
from the least action principle as Lagrange multipliers for
necessary constraints. Using these variables allows one to describe
the dynamics in terms of canonical (Hamiltonian) variables. The
nontrivial character of the Hamiltonian approach is due to the fact
that the fluid dynamics corresponds to the degenerated case,
see \cite{Dirac64,Gitman86}. 

Recently it was   shown \cite{KK_97,Kats_01} that the hydrodynamic
flows with discontinuities (including shocks) can be described in
terms of a least action principle, which includes (as well as
natural boundary conditions) the boundary conditions for the
subsidiary fields. In the present paper we show that all type of
discontinuities can be described by means of the least action
principle in terms of the canonical variables for the relativistic
flows.

\paragraph{Variational principle.}
The relativistic least action principle can be formulated in close
analogy to the nonrelativistic one. We introduce the action $A$,
\begin{equation}\label{26_04_3_0}
A = \int d^{4} x \, {\cal L} ,
\end{equation}
with the Lagrangian density
    \begin{equation}\label{23_02_04_4}
    \eqalign{
\mathcal{L} =  - \epsilon(n, S) + G J^{\alpha} Q_{, \alpha} \, ,
\\
G = (1, \nu_{B}, \Theta) , \q Q = (\varphi, \mu^{B}, S), \q B
= 1, 2, 3 , }
\end{equation}
where $\nu_{B}$, $\Theta$, $\varphi$, $\mu^{B}$ represent subsidiary
fields; $n$, $S$ and $\epsilon(n, S)$ denote the particle's number,
entropy and energy proper densities, $J^{\alpha} = n u^{\alpha}$ is
the particle current, and $u^{\alpha}$ is the four-velocity,
$u^{\alpha} = u^{0}(1, \mathbf{v}/c)$, $u^{0} = 1/\sqrt{1 -
\mathbf{v}^{2}/c^{2}}$; comma denotes partial derivatives. Small
Greek indexes run from $0$ to $3$, and the Latin indexes run from
$1$ to $3$; $x^{0} = ct$, $\mathbf{r} = (x^{1}, x^{2}, x^{3})$. The
metric tensor, $g^{\alpha \beta}$, corresponds to the flat
space-time in Cartesian coordinates, $g^{\alpha \beta} = \text{diag}
\{-1, 1,1,1\}$. The four-velocity obeys The normalization condition
\begin{equation}\label{23_04_03}
u^\alpha u_\alpha = g_{\alpha \beta} u^\alpha u^\beta = -1 .
  \end{equation}
Below we consider the four-velocity and the particle density $n$
as dependent variables expressed in terms of the particles current
$J^{\alpha}$,
    \begin{equation}\label{23_02_04_1}
u^{\alpha} =  J^{\alpha}/|J|  , \q n = |J| =  \sqrt{-
J^{\alpha}J_{\alpha}} \, .
\end{equation}
The fluid energy obeys the second thermodynamic law
\begin{equation}\label{24_04_03}
  d \epsilon = nT d S + n^{-1} w d n \equiv nT d S + W d n \, ,
\end{equation}
where $T$ is the temperature and $w \eq \epsilon + p$ is the
proper enthalpy density, $p$ is the fluid pressure, $W = w/n$.

Variation of the action given by  \equ{26_04_3_0} with respect to
the variables $J^{\alpha}$, $\Theta$, and $Q = (\varphi, \mu^{B},
S)$, which are supposed to be  independent, $A = A[J^{\alpha} ,
\varphi, \mu^{B} \, , S, \nu_{B} \, , \Theta]$,  results in the
following set of equations
   \begin{equation}\label{25_04_3_2}
\delta J^{\alpha} : \Longrightarrow W u_{\alpha} \eq V_{\alpha} =
- G Q_{,\alpha} \, ,
\end{equation}
    \begin{equation}\label{25_04_3_5}
\delta \varphi : \Longrightarrow J^{\alpha}_{\;\;, \alpha} = 0 ,
\end{equation}
    \begin{equation}\label{25_04_3_6}
\delta \mu^{B} : \Longrightarrow \partial_{\alpha} (J^{\alpha}
\nu_{B} ) = 0 , \q \text{or} \q D \nu_{A} =0,
\end{equation}
    \begin{equation}\label{25_04_3_7}
\delta \nu_{B}  : \Longrightarrow D \mu^{B} = 0 ,
\end{equation}
    \begin{equation}\label{25_04_3_3}
\delta S : \Longrightarrow  \partial_{\alpha} ( J^{\alpha} \Theta)
, \q \text{or} \q  D \Theta = - T ,
\end{equation}
    \begin{equation}\label{25_04_3_8}
\delta \Theta  : \Longrightarrow D S = 0 ,
\end{equation}
where $D \eq u^{\alpha} \p_{\alpha}$. \equ{25_04_3_2} gives us the
Clebsch type velocity representation, cf. Ref. \cite{Brown93}.
Contracting it with $u^{\alpha}$ results in the dynamic equation for
the scalar potential $\varphi$,
\begin{equation}\label{28_03_04}
D \varphi = W .
\end{equation}
Both triplets $\mu^{B}$ and $\nu_{B}$ represent the advected
subsidiary fields and do not enter the internal energy. Therefore,
it is natural to treat one of them, say, $\mu^{B}$ as the flow line
label.

Taking into account that the entropy and particle conservation are
incorporated into the set of variational equations, it is easy to
make sure that the equations of motion for the subsidiary variables
along with the velocity representation reproduces the relativistic
Euler equation. The latter corresponds to the orthogonal to the flow
lines projection of the fluid stress-energy-momentum $T^{\alpha
\beta}$ conservation, cf. Ref. \ct{Mizner-Torn-Wheeler73,LL_73},
\begin{equation}\label{1}
T^{\alpha \beta}_{\;\;\;\;, \beta} = 0 , \q T^{\alpha \beta} \eq
wu^\alpha u^\beta + p g^{\alpha \beta} .
\end{equation}
We may then write the relativistic Euler equation as
\begin{equation}\label{26_04_3_2}
( V_{\alpha , \beta} - V_{\beta, \alpha } ) u^{\beta} = T   S_{ ,
\alpha} \, ,
\end{equation}
where the thermodynamic relation
\begin{equation}\label{28_03_04_2}
d p = n d W - n T d S
\end{equation}
is taken into account. The vector $V_{\alpha}$, sometimes called the
Taub current, \cite{Taub59}, plays an important role in relativistic
fluid dynamics, especially in the description of circulation and
vorticity. Note that $W$ can be interpreted as an injection energy
(or chemical potential), cf., for instance
\cite{Mizner-Torn-Wheeler73}, i.e., the energy per particle required
to inject a small amount of fluid into a fluid sample, keeping the
sample volume and the entropy per particle $S$ constant. Therefore,
$V_{\alpha}$ is identified with the four-momentum per particle of a
small amount of fluid to be injected in a larger sample of fluid
without changing the total fluid volume and the entropy per
particle.

\paragraph{Boundary conditions.}
In order to complete the variational approach for the flows with
discontinuities, it is necessary to formulate the boundary
conditions for the subsidiary variables, which do not imply any
restrictions on the physically possible discontinuities (the shocks,
tangential and contact discontinuities), are consistent  with the
corresponding dynamic equations, and thus are equivalent to the
conventional boundary conditions, i.e., to continuity of the
particle and energy-momentum fluxes intersecting the discontinuity
surface $R(x^{\alpha}) = 0$, cf. Ref. \cite{LL_73},
\begin{equation}\label{29_03_04}
\{ \breve{J}\} = 0 , \q \breve{J} \eq J^{\alpha} n_{\alpha} \, ,
\end{equation}
\begin{equation}\label{29_03_04_1}
\{ T^{\alpha \beta} n_{\beta}\} = 0 ,
\end{equation}
where $n_{\alpha} $ denotes the unit normal vector to the
discontinuity surface,
\begin{equation}\label{29_03_04_2}
n_{\alpha} = N_{\alpha}/ N  , \q N_{\alpha} = R_{, \alpha} \, \q N
= \sqrt{N_{\alpha} N^{\alpha} } \, ,
\end{equation}
and braces denote jump, $\{ X\} \eq X|_{R = +0} - X|_{R = -0}$.

Our aim is to obtain boundary conditions as natural boundary
conditions for the variational principle. In the process of deriving
the volume equations we have applied integration by parts to the
term $J^{\alpha} G  \delta Q_{, \alpha}$. Vanishing of the
corresponding surface term along with that resulting from the
variation of the surface itself will lead to the appropriate
boundary conditions after the variational principle has been
specified.

Rewriting the (volume) action with the discontinuity surface being
taken into account in  explicit form as
\begin{equation}\label{27_04_03_B6}
A = \int d^{4} x \sum_{\varsigma = \pm 1} \mathcal{L}^{\varsigma}
\theta (\varsigma R) \, ,
\end{equation}
where 
$\theta $ stands for the step-function, we obtain the residual
part of the (volume) action in the form
\begin{equation}\label{27_04_03_B6A}
\left. \delta A \right|_{res} = \int d^{4} x \sum_{\varsigma = \pm
1} \left[ \varsigma \mathcal{L} \delta_{D} ( R) \delta R +
\theta(\varsigma R) \p_{\alpha}  ( J^{\alpha} G  \delta Q )
\right] .
\end{equation}
Here $\delta_{D}$ denotes Dirac's delta-function and we omit the
index $\varsigma$ labeling the quantities that correspond to the
fluid regions divided by the interface at $R = 0$; the superscript
$\varsigma \gtrless 0$ corresponds to the quantities in the regions
$R \gtrless 0$, respectively. Integrating the second term by parts
and supposing that the surface integral
    $
\int d^{4} x \sum_{\varsigma = \pm 1} \p_{\alpha}
\left(\theta(\varsigma R)  (u^{\alpha} G  \delta Q ) \right)
    $
vanishes due to vanishing of the variations $\delta Q$ at
infinity, we arrive at the residual action expressed by the
surface integral
\begin{equation}\label{30_04_3_10}
\left. \delta A \right|_{res} = \int d^{4} x \sum_{\varsigma = \pm
1} \varsigma \delta_{D} ( R) \left[ \mathcal{L} \delta R - R_{,
\alpha} J^{\alpha} G  \widetilde{\delta} Q \right] .
\end{equation}
here $\widetilde{\delta} Q$  designates the limit values of the
volume variations, $\widetilde{\delta} Q^{\pm} \eq (\delta Q)_{R =
\pm 0}$. It is convenient to express  these variations in terms of
variations of the boundary restrictions of the volume variables,
$\delta (X_{R = \pm 0}) \eq \delta\widetilde{X}^{\pm}$, and
variation of the discontinuity surface. It is easy to show that
\begin{equation}\label{1_05_3_7}
\widetilde{\delta} X = \delta \widetilde{X} + |N|^{-1} n^{ \alpha}
X_ {, \alpha} \delta R - X_ {, \alpha} P^{\alpha}_{\;\; \beta}
\delta f^{\beta} \,  ,
\end{equation}
where $P^{\alpha}_{\;\; \beta} = \delta^{\alpha}_{\;\; \beta} -
n^{\alpha}n_{ \beta}$, and $\delta f^{\beta}$ is an arbitrary
infinitesimal four-vector related to the one-to-one mapping of the
surfaces $R = 0$ and $R + \delta R = 0$.

Vanishing  of the action variation with respect to variations of
the surface variables $\delta R$ and $\delta f^{\beta}$ (which are
supposed to be independent) results in the following boundary
conditions
\begin{equation}\label{28_05_03}
\delta R : \Rightarrow   \left\{  p  + (u^{\alpha} n_{\alpha})^{2}
w \right\} = 0 ,
\end{equation}
\begin{equation}\label{28_05_03_1}
\delta f^{\beta} :  \Rightarrow P^{\gamma}_{\;\;\beta}  \left\{ W
J^{\alpha} N_{\alpha} u_{\gamma} \right\} = 0 , \;\; \mathrm{or}
\;\; P^{\gamma}_{\;\;\beta} \left\{ \check{J}  W  u_{\gamma}
\right\} = 0  ,
\end{equation}
which are equivalent to continuity of the momentum and energy
fluxes, cf. \equ{29_03_04_1}. Here we consider that the `on shell'
value of the volume Lagrangian density, $\mathcal{L}_{eq}$, is
equal to the pressure, $\mathcal{L}_{eq} = -\epsilon + n G D Q = -
\epsilon + w = p$.

Now we can complete formulation of the variational principle
appropriate both for continuous and discontinuous flows. The
independent volume variables are indicated above, and independent
variations of the surface variables are $\delta R$, $\delta
f^{\beta}$,  the variations of the surface restrictions of the
generalized coordinates $\delta \varphi$, $\delta \mu^{B}$, supposed
to be equal from both sides of the discontinuity, $\{ \delta \varphi
\} = \{ \delta \mu^{B} \} =0$, and $\delta S$ with $\{ \delta S \}
\ne 0$. Under these assumptions we arrive at the following subset of
the boundary conditions
\begin{equation}\label{28_05_03_2B}
\delta \widetilde{\varphi}: \Rightarrow    \{ J^{\alpha}
n_{\alpha}  \} \eq \{ \check{J} \}= 0  \quad \text{for} \quad \{
\delta \widetilde{\varphi} \} = 0 ,
\end{equation}
\begin{equation}\label{28_05_03_2C}
\delta \widetilde{\mu}^{B} : \Rightarrow   \{ \nu_{B} J^{\alpha}
n_{\alpha}  \} \eq \check{J} \{ \nu_{B} \} = 0  \quad \text{for}
\quad \{ \delta \widetilde{\mu}^{B} \} = 0 ,
\end{equation}
\begin{equation}\label{28_05_03_2}
\delta \widetilde{S}^{\pm} : \Rightarrow J^{\alpha}
n_{\alpha}\widetilde{ \Theta}^{\pm} \eq \check{J} \widetilde{
\Theta}^{\pm} = 0 .
\end{equation}
Eqs.~\eqref{28_05_03}--\eqref{28_05_03_2B} reproduce the usual
boundary conditions, and Eqs.~\eqref{28_05_03_2C},
\eqref{28_05_03_2} are the boundary conditions for the subsidiary
variables. Other boundary conditions for the latter variables do not
strictly follow from the variational principle under discussion. But
we can find them from the corresponding volume equations of motion,
providing, for instance, that they are as continuous as
possible.\footnote{Note that the choice of the boundary conditions
for the fields $\varphi$, $\mu^{B}$, $\nu_{B}$ and $\Theta$ is not
unique due to the fact that they play the roles as generalized
potentials and therefore possess the corresponding gauge freedom
relating to the transformations $\varphi, \mu^{B}, \nu_{B}, \Theta
\rightarrow \varphi', \mu'^{B}, \nu'_{B}, \Theta'$ such that
$u'_{\alpha} = u_{\alpha}$ (given by the representation
\eqref{25_04_3_2}). For instance, it seems possible to use entropy
$S$ as one of the flow line markers. But if we are dealing with
discontinuous flows then it is necessary to distinguish the Lagrange
markers of the fluid lines, $\mu^{B}$, and the entropy, $S$. Namely,
the label of the  particle intersecting a shock surface evidently
does not change, but the entropy does change. Thus, entropy can be
chosen as one of the flow line markers only for the flows without
entropy discontinuities.} The natural choice corresponds to
continuity of their fluxes,
\begin{equation}\label{27_04_03_C13}
  \{  n_{\alpha} u^{\alpha}  n \mu^{B}  \} \eq \check{J} \{ \mu^{B}  \} =0  ,
\end{equation}
\begin{equation}\label{27_04_03_C14}
   \{ n_{\alpha} u^{\alpha}  n \varphi  \} \eq \check{J} \{ \varphi  \} = 0 .
\end{equation}
The set of the boundary conditions given by
Eqs.~\eqref{28_05_03}--\eqref{27_04_03_C14} is complete and allows
one to describe any type of discontinuities, including shocks. For
the latter case $\check{J} \ne 0$ and we arrive at continuity of the
variables $\nu_{B}$, $\mu^{B}$, $\varphi$ and zero boundary value of
$\Theta$. For $\check{J} = 0$ the flow lines do not intersect the
discontinuity surface and we obtain very weak restrictions on the
boundary values of the subsidiary variables, cf. the nonrelativistic
case discussed in Refs.~\cite{KK_97,Kats_01}. Note that for the
specific case $\check{J} = 0$ (slide and contact discontinuities) we
can simplify the variational principle assuming all both-side
variations of the subsidiary variables to be independent.

The above variational principle allows modifications. First, it is
possible to exclude constraints, expressing the four-velocity by
means of the representation \eqref{25_04_3_2}. In this case the
volume Lagrangian density can be chosen to coincide with the fluid
pressure, cf. Ref. \cite{Brown93}, where the continuous flows are
discussed in detail. Second, we can include into the action the
surface term respective for the surface constraints, cf.
Refs.~\ct{KK_97,Kats_01,KATS_02,KATS_02A}, where such surface terms
are discussed for ideal hydrodynamics and magnetohydrodynamics in
the nonrelativistic limit. This can be done for the cases both with
excluded and non excluded volume constraints.

{\textbf{Canonical variables.}} Starting from the action in
\equ{26_04_3_0} and Lagrangian density given by \equ{23_02_04_4}
we can introduce the canonical (Hamiltonian) variables according
to the general receipt. Let $Q$ represents the canonical
coordinates then
\begin{equation}\label{2_02_04}
P \eq \frac{\delta A}{\delta Q_{,0}} = J^{0} G \eq (\pi_{\varphi}
\, , \pi_{\mu^{B}}, \pi_{S})
\end{equation}
provides the conjugate momenta. Relations \eqref{2_02_04} cannot be
solved for the generalized velocities $Q_{,0}$ suggesting that we
are dealing with the degenerated (constraint) system, cf. Refs.
\cite{Dirac64,Gitman86,GP93,ZakharovKuznetsov97}. But the
constraints are of the first type. Thus,  performing the Legendre
transform with respect to $Q$ we arrive at the Hamiltonian density
\begin{equation}\label{2_02_04_2}
\mathcal{H} =  P Q_{,0} - p(W,S)  ,
\end{equation}
where we suppose that the four-velocity is given by representation
\eqref{25_04_3_2}. Making use of the definition \eqref{2_02_04}
and of the time component of the velocity representation,
\equ{25_04_3_2}, we can transform the first term in
\equ{2_02_04_2} as
\begin{equation}\label{2_02_04_2B}
P Q_{,0} = J^{0} G Q_{,0} = - \pi_{\varphi} V_{0} = \pi_{\varphi}
V^{0} .
\end{equation}
Taking into account the normalization condition for the Taub
current, $V_{\alpha}V^{\alpha} = - W^{2}$, we obtain
\begin{equation}\label{2_02_04_2C}
V^{0} = \sqrt{W^{2} + V_{a}V^{a}} \, .
\end{equation}
Consequently, we arrive at the following Hamiltonian density
\begin{equation}\label{2_02_04_2A}
\mathcal{H} \eq \mathcal{H}(P,Q, Q_{, a}; W) = \sqrt{W^{2} +
V_{a}V^{a}} \, \pi_{\varphi} - p(W, S) .
\end{equation}
In terms of the canonical coordinates and momenta the space
components of the velocity are
\begin{equation}\label{24_05_04}
\pi_{\varphi} V_{a} = - P Q_{,a} .
\end{equation}
The canonical equations following from this Hamiltonian reproduce in
a $3+1$ form the  dynamical equations above for the variables
entering the Taub current representation. Variation of the action
with respect to the chemical potential $W$ results in the identity
\begin{equation}\label{22_02_04_4}
n = \frac{\pi_{\varphi}}{\sqrt{1 + V_{a}V^{a}/W^{2}}}   \, .
\end{equation}
Obviously, this relation is equivalent to \equ{2_02_04_2C},
expressing the particle density $n$ in terms of the variables
entering the Hamiltonian.

We emphasize that the Hamiltonian given by \equ{2_02_04_2A} depends
not only on the generalized coordinates $\varphi$, $\mu^{B}$, $S$,
their spatial derivatives and conjugate momenta, but also on the
chemical potential $W$ as well.   Evidently, we can consider $W$ as
the additional generalized coordinate with zero conjugate momentum,
$\pi_{W} = 0$. This condition is consistent with the dynamic
equations due to the fact that $\p_{0} \pi_{W} = \p \mathcal{H}/ \p
W = 0$, cf. \equ{22_02_04_4}.

Bearing in mind that we are dealing with flows having
discontinuities, it is seen that in the discussed variant of the
least action principle we do not arrive at the additional surface
variables except for the variable defining the discontinuity
surface, $R$. But this enters the action functional without
derivatives. Therefore, the corresponding conjugate momentum is
zero-valued. Introducing the Hamiltonian variables for the flows
with discontinuities we have to treat $R$ as the surface function,
defining some (surface) constraint. The latter is nothing else than
continuity of the normal component of the fluid momentum flux,
\equ{28_05_03}.

%

\paragraph{Local invariants and recursion relations.}
In addition to energy, momentum, and angular momentum conservation,
for the ideal hydrodynamic flows there exist specific local
conservation laws related to the advected and frozen-in fields, and
corresponding  topological invariants (vorticity, helicity, Ertel
invariant, etc.), cf. Refs.
\cite{lamb,Salmon82,STY_90,GP93,ZakharovKuznetsov97} and citations
therein for the nonrelativistic case. They are caused by the
relabeling symmetry, cf. Ref. \cite{Salmon82}. Discussion of these
problems along with the recursion relations linking the four
different types of invariants  for the relativistic flows seems
insufficient or absent in the literature , see Refs.
\cite{Taub59,Schutz70,Brown93,ZakharovKuznetsov97,KATZ_84} and
citations therein. Exploitation of the above description permit us
considering these invariants in a simplified form. Here we shall
briefly discuss  the invariants and recursion relations.

The local Lagrangian invariants, say $I$, correspond to advected
(dragged) quantities,
\begin{equation}\label{25_06_04}
D I = 0.
\end{equation}
The partial derivative of each scalar Lagrange invariant gives us
the simplest example of the Lamb type momentum, $L_{ \alpha}$,
which satisfy the following relations
\begin{equation}\label{29_10_3_4}
D L_{ \alpha}  +  u^{\beta}_{ \; , \alpha} L_{ \beta} = 0  , 
\end{equation}
The next type of invariants are vector
conserved quantities, $X^{ \alpha}$, being proportional to the
four-velocity, i.e.,
\begin{equation}\label{25_06_04_1}
X^{ \alpha}_{ \; , \alpha} = 0, \q X^{ \alpha} =  |X| u^{ \alpha}
.
\end{equation}
The trivial example of such quantities is the particle current $J^{
\alpha} = n u^{ \alpha}$. The last type corresponds to the frozen-in
fields, $M_{ \alpha \beta} $, defined as the
antisymmetric tensors obeying the following equation 
\begin{equation}\label{30_10_03_2}
D M_{ \alpha \beta} + u^{\gamma}_{ \; , \alpha} M_{ \gamma \beta}
+ u^{\gamma}_{ \; , \beta} M_{ \alpha \gamma} = 0 , \q M_{ \beta
\alpha} = - M_{ \alpha \beta}  \, .
\end{equation}

Now we can derive some recursion relations. First, an arbitrary
function of the Lagrangian invariants is also Lagrangian invariant,
\begin{equation}\label{29_10_3_8B}
I'' = F(I, I', \ldots)  \, .
\end{equation}
Second, multiplication of any invariant by the Lagrange invariant
results in the invariant of the same type. Symbolically,
\begin{equation}\label{25_06_04_2}
L'_{ \alpha} = I L_{ \alpha}  \, , \q X'^{ \alpha} = I X^{ \alpha}
\, , \q M'_{ \alpha \beta}  = I M_{ \alpha \beta} .
\end{equation}
Third,
\begin{equation}\label{26_06_04}
L_{ \alpha} = I_{, \alpha} \, ,
\end{equation}
\begin{equation}\label{29_10_3_5}
M_{ \alpha \beta}  =    L_{ \alpha, \beta} - L_{ \beta, \alpha} \,
,
\end{equation}
\begin{equation}\label{7_03_04_D4_1}
M_{\alpha \beta} = L_{\alpha} L'_{ \beta} - L_{\beta} L'_{ \alpha}
\, ,
\end{equation}
\begin{equation}\label{28_06_04}
L_{ \alpha} = M_{ \alpha \beta} u^{ \beta}  \, ,
\end{equation}
\begin{equation}\label{5_11_03_A}
X^{\alpha} = \epsilon^{\alpha  \beta \mu \nu } I_{, \beta } I'_{
,\mu } I''_{ , \nu } \, ,
\end{equation}
\begin{equation}\label{8_03_04_1_A}
I''' = n^{-1} \epsilon^{\alpha \beta \mu \nu } u_{\alpha} I_{,
\beta } I'_{ , \mu } I''_{ , \nu } \, , \q D I''' = 0 .
\end{equation}
Here $\epsilon^{\alpha \beta \mu \nu }$ is Levi-Civita tensor. Note
that the latter relation follows from the fact that the conserved
current defined by \equ{5_11_03_A} is collinear to the
four-velocity, $X^{\alpha} = |X| u^{\alpha} \eq - |X| J^{\alpha}/n$.
Consequently, $|X| = - X^{\alpha} u_{\alpha} $, and $|X|/n  =  I $
is an advected scalar. Note also that we can easily arrive at the
conserved four-currents, say $Z^{\alpha}$, which are not necessarily
collinear to the velocity field. In analogy with relation
\eqref{5_11_03_A} we find
\begin{equation}\label{26_06_04_1}
Z^{\alpha}_{\; \;, \alpha} = 0 \q \mathrm{for}  \q  Z^{\alpha} =
\epsilon^{\alpha \beta \mu \nu } L_{ \beta ,\mu } I_{ , \nu } \, .
\end{equation}
Other examples of such conserved currents are as follows
\begin{equation}\label{28_06_04_1}
Z^{\alpha} = \epsilon^{\alpha \beta \mu \nu }  \left( I L_{ \beta,
\mu} \right)_{, \nu}  \, ,
\end{equation}
\begin{equation}\label{28_06_04_2}
Z^{\alpha} = \epsilon^{\alpha \beta \mu \nu } \left( I L_{ \beta}
L'_{\mu} \right)_{, \nu} \, ,
\end{equation}
\begin{equation}\label{28_06_04_3}
Z^{\alpha} = \left( {^{\ast} \hspace{-0,2em}M}^{\alpha  \beta } I
\right)_{, \beta}  \, ,
\end{equation}
where ${^{\ast} \hspace{-0,2em}M}^{\alpha  \beta }$ is dual to
$M_{\alpha  \beta }$.

These recursion relations represent a strict analog of the
nonrelativistic ones, cf. Refs.
\cite{GP93,STY_90,ZakharovKuznetsov97} and can be proved by direct
calculations. Existence of these four types of invariants is related
to the dimensionality of the space-time as becomes evident if we
remember that they correspond to the differential forms of $0$th,
$1$st, $2$nd and $3$rd order, cf., for instance, Ref.
\cite{Mizner-Torn-Wheeler73}. The recursion procedure allows one to
study the structure of the invariants.

It is noteworthy that not all invariants could be obtained by means
of the recursion procedure. For flows of general type there also
exists the Ertel invariant (or the potential vorticity), cf. Ref.
\cite{KATZ_84}. The corresponding conserved four-current is of the
form
\begin{equation}\label{9_11_03}
\mathcal{E}^{\alpha} = - \frac{1}{2} \epsilon^{\alpha  \beta \mu
\nu } \omega_{ \beta  \mu } S_{ ,\nu } =  - {^{\ast}
\hspace{-0,2em}\omega}^{\alpha  \nu } S_{ ,\nu }  \, ,
\end{equation}
where $\omega_{ \beta \mu }$ is the (Khalatnikov) vorticity
tensor, $\omega_{ \beta  \mu } = V_{\mu , \beta} - V_{\beta ,
\mu}$, and ${^{\ast} \hspace{-0,2em}\omega}^{\alpha \nu }$ is its
dual. The vorticity tensor obeys the following equation
\begin{equation}\label{14_01_03_7A}
D \omega_{\alpha \nu } + \omega_{\alpha \beta}u^{\beta}_{\;\;,
\nu} + \omega_{\beta \nu}u^{\beta}_{\;\;, \alpha} = T_{, \nu} S_{,
\alpha} - T_{, \alpha} S_{, \nu}  \eq n^{-2} (n_{, \alpha} p_{,
\nu} - n_{, \nu} p_{, \alpha})  \, .
\end{equation}
It can be proved that $\mathcal{E}^{\alpha}$ is divergence-free
(the easiest way is to use the above velocity representation) and
\begin{equation}\label{15_02_04}
\mathcal{E}^{\alpha} = \mathbb{E} J^{\alpha} , \q D \mathbb{E} =
0,
\end{equation}
where $\mathbb{E} $ is a direct generalization of the well-known
nonrelativistic potential vorticity. If the vorticity tensor would
be the frozen-in quantity then conservation of the vector given by
\equ{9_11_03} would follow from the above recursion relations. But
for the non-barotropic flows 
it is not so. Here $\omega_{\alpha \nu }$  obeys \equ{14_01_03_7A}
and becomes a frozen-in field only for  the barotropic (isentropic,
in particular) flows. Nevertheless, the Ertel current is conserved.
Therefore, the basic invariants may be obtained by direct
calculations.

It is interesting to note that for the non-barotropic flows there
exists a conserved current generalizing the helicity current.
Consider the reduced Taub vector,
\begin{equation}\label{26_01_04}
\widetilde{V}_{\alpha}  \eq V_{\alpha} + \Theta S_{, \alpha}  \, ,
\end{equation}
where $\Theta $ obeys \equ{25_04_3_3}, and the corresponding
reduced vorticity tensor
\begin{equation}\label{26_01_04_1}
\widetilde{\omega}_{ \alpha \beta} \eq \widetilde{V}_{\beta ,
\alpha} - \widetilde{V}_{\alpha , \beta} = \nu_{A, \beta}
\mu^{A}_{, \alpha} - \nu_{A, \alpha} \mu^{A}_{, \beta} \, .
\end{equation}
This tensor is orthogonal to the flow lines,
\begin{equation}\label{26_01_04_5}
\widetilde{\omega}_{ \alpha \beta} u^{\beta} = 0,
\end{equation}
therefore, the reduced helicity current
\begin{equation}\label{26_01_04_7}
\widetilde{Z}^{\alpha}  = {^{\ast}
\hspace{-0,1em}\widetilde{\omega}}^{\alpha \nu } \widetilde{V}_{
\nu }
\end{equation}
is conserved for arbitrary flows,
\begin{equation}\label{26_01_04_8}
\widetilde{Z}^{\alpha}_{\;\; , \alpha}  = {^{\ast}
\hspace{-0,1em}\widetilde{\omega}}^{\alpha \nu }  \widetilde{V}_{
\nu , \alpha}  = \frac{1}{2} {^{\ast}
\hspace{-0,1em}\widetilde{\omega}}^{\alpha \nu }
\widetilde{\omega}_{\alpha \nu  } \eq \frac{1}{4} \epsilon^{\alpha
\nu \beta \gamma } \widetilde{\omega}_{\beta \gamma  }
\widetilde{\omega}_{\alpha \nu  }  = 0  .
\end{equation}
Here the ``thermassy'' field $\Theta$ can be chosen in such a way
that its initial value is zero and thus the initial value of the
generalized helicity coincides with the conventional one.

To derive a set of  local invariants we have to start with those
following directly from the hydrodynamic equations and then apply
the above recursion relations. As the simplest example consider the
non-barotropic flow. Then one can start with the specific entropy
$S$ and the Ertel invariant $\mathbb{E}$. It is easy to show that
the general form of the gauge-independent scalar invariants is
\begin{equation}\label{28_06_04_4}
I = F(S, \mathbb{E}),
\end{equation}
where $F$ is arbitrary function, cf. the nonrelativistic case,
\cite{ZakharovKuznetsov97}. The structure of the complete set of
invariants differs for the different type of flows and will be
discussed in forthcoming publications. For instance, for the
barotropic flows we have the independent frozen in field $M_{
\alpha \beta} = \omega_{ \alpha \beta}$ in addition to the scalar
invariants of the first generation, $S$ and $\mathbb{E}$. This
fact allows one to obtain a more complicated set of the
invariants.


\paragraph{Conclusion.}

We have shown that it is possible to describe the relativistic ideal
fluids with all physically allowable discontinuities in terms of the
least action principle both in the Lagrangian and Hamiltonian
description. The boundary conditions for the subsidiary variables,
entering the Clebsch type velocity representation, are obtained in
two different ways: one way follows from the variational principle
as natural boundary conditions while the other one was obtained from
the dynamical equations under the assumption relating to the absence
of the corresponding sources and the maximal continuity compatible
with the volume equations. It is possible to change the variational
principle in such a way that all boundary conditions will result
from it, i.e., they become natural boundary conditions. For this
purpose it is necessary to modify the variational principle by
adding a surface term with corresponding constraints, similarly to
the nonrelativistic case (compare with the papers
\cite{KK_97,Kats_01} for the hydrodynamics and
\cite{KATS_02,KATS_02A} for the magnetohydrodynamics). These
variants will be discussed in future works.

The approach discussed in this paper allowed us to give a simple
treatment of the additional invariants of  motion and present a set
of recursion relations linking different types of invariants. In
particular, we presented a generalization of the helicity invariant
for the non-barotropic relativistic flows. This approach is suitable
for the general relativity and for the relativistic
magnetohydrodynamics as well. 
The discontinuous flows for the general relativity can be described
in analogy with the above discussion and the results will be
published elsewhere.

\subsection*{Acknowledgment}

\frenchspacing

This work was supported by INTAS (Grant No. 00-00292).

\end{document}